\documentclass[aps,prb,reprint,preprintnumbers]{revtex4-1}
\usepackage{graphicx,bm,epsfig,ulem,color,mathrsfs,dcolumn,multirow,setspace}
\usepackage{array,amsmath,amssymb,dsfont,gensymb}
\usepackage{bibentry,natbib}
\usepackage{booktabs}
\usepackage{color}
\usepackage{enumitem}
\usepackage[utf8]{inputenc}
\usepackage[english]{babel}
\newcolumntype{M}[1]{>{\centering\arraybackslash}m{#1}}
\newcolumntype{N}{@{}m{0pt}@{}}
\usepackage{lineno}

\begin{document}
\bibliographystyle{unsrt}
\title{On the Sign of Fermion-Mediated Interactions}

\author{Qing-Dong \surname{Jiang}}

\affiliation{Department of Physics, Stockholm University, Stockholm SE-106 91 Sweden}
\begin{abstract}
We develop a unified framework for understanding the sign of fermion-mediated interactions by exploiting the symmetry classification of Green's functions. In particular, we establish a theorem regarding the sign of fermion-mediated interactions in systems with chiral symmetry. The strength of the theorem is demonstrated within multiple examples with an emphasis on electron-mediated interactions in materials.\end{abstract}
\preprint{}
\maketitle

{\it Introduction.---}That the exchange of a particle can produce a force is one of the most remarkable conceptual advances in physics. Each of the fundamental interactions has an associate bosonic force carrier: for example, photons mediate Coulomb interactions and gravitons mediate gravitational interactions. 
Given these boson-mediated interactions, one may naturally ask the following innocuous question: ``Can fermions also mediate interactions?" 

While the answer is `yes', there is an essential difference between boson-mediated interactions and fermion-mediated interactions. That is, due to the conservation of fermionic parity, fermions need to be exchanged at least twice to produce a force whereas bosons only need to be exchanged once (see the Feynman diagram in Fig. 1). As this Feynman diagram resembles that of the Casimir effect, fermion-mediated interactions are occasionally called fermionic Casimir effects in the literature\cite{fermCa0}. In the case of one-boson-mediated interactions, the sign is uniquely determined by the spin of the exchanged particles:  Exchanging a scalar, or a tensor particle produces a universally attractive force, while exchanging a vector particle can produce a repulsive force between like charges \cite{azee}. However, unlike the bosonic case, a universal understanding of the sign of fermion-mediated interactions is currently lacking. 

In this work, we study the sign of various fermion-mediated interactions. In condensed-matter physics, electron-mediated interactions were initially proposed to explain the ordering of adsorbates at surfaces\cite{schrieffer,repp}, and were recently considered to be crucial for engineering the properties of novel materials such as graphene\cite{cheia}. As an important mechanism for magnetic ordering, the Ruderman–Kittel–Kasuya–Yosida (RKKY) interaction is another example of electron-mediated interactions\cite{rkkyref}. In the cold-atom area, fermion-mediated interactions have received extensive theoretical investigations\cite{fmicoldatomth} and were recently observed in experiments consisting of a mixture of bosonic and fermionic quantum gases\cite{fmicoldatom}. In high-energy physics, fermion-mediated interactions are of particular relevance to the physics of neutron stars\cite{fermCaneutron} and quark-gluon plasma\cite{fermCaquark1}. 
Given the ubiquitous presence of fermion-mediated interactions, their sign is of both theoretical interest and practical significance\cite{levitov,lebo,signfmi}. 
For example, attractive fermion-mediated interactions could lead to new phases of matter such as supersolids\cite{blatter}.
\begin{figure}[!htb]
\includegraphics[height=3.6cm, width=6. cm, angle=0]{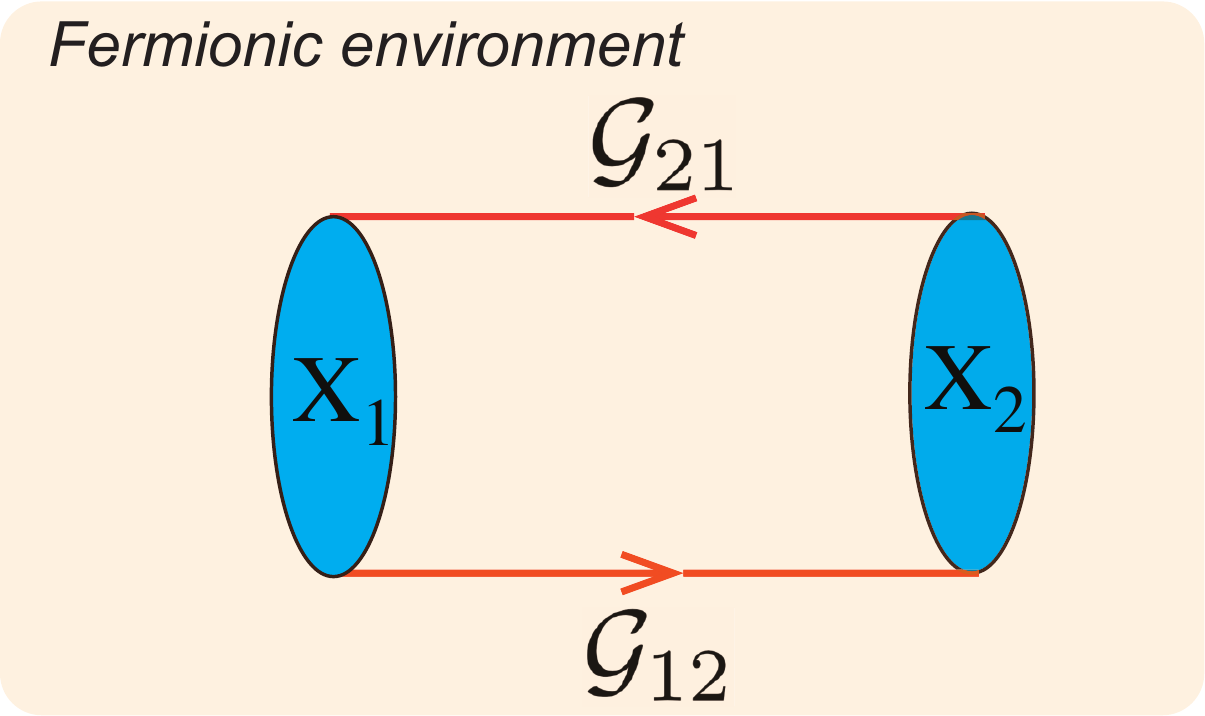}
\caption{Feynman diagram for fermion-mediated interactions.  The red lines represent fermionic Green's functions (propagators) $\mathcal G_{12}$ and $\mathcal G_{21}$ that connect $\rm X_{1}$ and $\rm X_{2}$. Since two Green's functions are involved in the Feynman diagram, we say that objects $\rm X_1$ and $\rm X_2$ interact with each other by exchanging fermions twice. The blue bubbles represent the scattering matrices of $\rm X_1$ and $\rm X_2$, the form of which will be given in the text. \label{fig1}}
\end{figure}

By exploiting the symmetry classification of Green's functions\cite{ludwig,gurarie},  we present a unified framework for understanding the sign of fermion-mediated interactions, eliminating some loopholes and resolving some controversies in the literature. The symmetry that plays the central role is called chiral symmetry, which is a combination of time-reversal symmetry and charge-conjugation symmetry. And the associated chirality corresponds to whether an operator is even or odd under the action of the chiral symmetry.
Specifically, in systems with chiral symmetry, we show that the sign of fermion-mediated interactions $\rm U_{12}$ between objects $\rm X_1$ and $\rm X_2$ is given by the simple rule:
\begin{eqnarray}\label{signu}
{\rm sign\left(U_{12}\right)}=(-1)^{\eta}\chi_1\chi_2.
\end{eqnarray}
Here,  $\chi_{1(2)}=\pm 1$ is the chirality of the object $\rm X_{1(2)}$; $\eta=0$ or $1$ depending on the strength of scattering potentials being strong or weak.
Between objects with the {\it same} chirality, a strong scattering potential leads to repulsion, whereas a weak scattering potential leads to attraction. By sharp contrast, between objects with {\it opposite} chiralities, a strong scattering potential leads to attraction, whereas a weak scattering potential leads to repulsion. 

The rest of the paper is organized as follows. We first derive a nonperturbative expression of fermion-mediated interactions involving Matsubara Green's functions and scattering matrices. Then, we show how nonspatial (local) symmetries constrain the Green's functions and scattering matrices of a general interacting system. With the above preparation, we establish a theorem regarding the sign of fermion-mediated interactions. Finally, we demonstrate the power of our theorem by giving a number of examples with an emphasis on electron-mediated interactions in materials.

{\it Field theory of fermion-mediated interactions.---}To set the stage, we now present a unified expression for fermion-mediated interactions. While various expressions for fermion-mediated interactions exist in the literature\cite{levitov,lebo,fmiexpress}, a universal, and nonperturbative derivation is not entirely trivial. A related path-integral approach has been used to study the Casimir effect in Refs. \onlinecite{keneeth,jiang}.

Consider two objects ${\rm X}_i$ ($i=1,2$) embedded at the positions $\bold x_i$ in a fermionic environment, and introduce localized potential operators $\mathcal{\hat V}_i$ to describe the scattering effect of ${\rm X}_i$. We consider bilinear coupling $\bar \psi \mathcal{\hat V}_i\psi$ between the operators $\mathcal{\hat V}_{i}$ and the mediating fermionic fields ($\bar \psi, \psi$); thus the effective Euclidean action reads (we use units $\hbar=c=1$)
 \begin{eqnarray}\label{actions}
 \mathcal S=-\sum_{n} \int d\bold x~ \bar \psi_n(\bold x) \left(\mathcal{\hat G}^{-1}-\mathcal{\hat V}_1-\mathcal{\hat V}_2\right )\psi_n(\bold x),
 \end{eqnarray}
where $\mathcal{\hat G}=(i\omega_n+\hat H)^{-1}$ represents the Matsubara Green's function of the fermionic host.  The partition function of the whole system is
$ \mathcal Z= \int \mathcal D\bar\psi \mathcal D\psi~\exp{\left(-\mathcal S\right)}$. In the absence of ${\rm X}_i$, the partition function $\mathcal Z_0$ can be obtained from $\mathcal Z$ by setting $\mathcal{\hat V}_i=0$. Consequently, the change of the energy due to the introduction of ${\rm X}_i$  can be formally obtained from the reduced partition function
\begin{eqnarray}
\mathcal E&=&- \frac{1}{\beta}\ln \frac{\mathcal Z}{\mathcal Z_0}\nonumber\\
  &=&-\frac{1}{\beta} \ln{\rm det} \left[1-\mathcal{G}(\bold x,\bold x^\prime)\,\mathcal{V}_1-\mathcal{ G}(\bold x,\bold x^\prime)\,\mathcal{ V}_2\right],
\end{eqnarray}
where $\mathcal G(\bold x,\bold x^\prime )=\langle \bold x|\mathcal{\hat G}|\bold x^\prime \rangle$, $\mathcal V_i=\langle \bold x^\prime|\mathcal{\hat V}_i |\bold x^\prime\rangle \,\delta_{\bold x^\prime,\bold x_i}$, and $\beta$ is the inverse temperature. 
Note that the total energy $\mathcal E$ contains three parts: the self energies of ${\rm X}_{1}$ and ${\rm X}_{2}$ and the mutual interaction between them. Therefore, to obtain the interaction energy, we need to subtract the self-energy contribution that does not depend on relative positions of ${\rm X}_1$ and ${\rm X}_2$.
For this purpose, it is convenient to put the energy in a matrix form: 
 \begin{eqnarray}
\mathcal E=-\frac{1}{\beta} \ln {\rm det} \left(
 \begin{array}{cc}
 1-\mathcal{G}_{11}\mathcal V_1&-\mathcal G_{12} \mathcal V_2\\
-\mathcal G_{21} \mathcal V_1&1-\mathcal G_{22}\mathcal V_2
 \end{array}\right).
 \end{eqnarray}
Here, $\mathcal G_{12}\equiv\mathcal G(i\omega_n, \bold r)$ and $\mathcal G_{21}\equiv\mathcal G(i\omega_n,-\bold r)$ are the Matsubara Green's functions linking ${\rm X}_1$ with ${\rm X}_2$;  $\mathcal G_{11}=\mathcal G_{22}\equiv \mathcal G(i\omega_n,0)$ are local Matsubara Green's functions; $\bold r=\bold x_2-\bold x_1$ denotes the relative distance between ${\rm X}_1$ and ${\rm X}_2$; and $\omega_n=(2n+1)\pi \beta^{-1}$ are fermionic Matsubara frequencies.
After subtracting the self-energies, i.e., the diagonal contribution of the matrix,  we obtain the universal formula for two-particle exchange interactions
 \begin{eqnarray}\label{uform}
 {\rm U_{12}}=
- \frac{1}{\beta} \sum_{n}\ln {\rm det} \left(1-\mathcal G_{12}\, T_2\,\mathcal G_{21}\,T_1\right),
 \end{eqnarray}
where $T_1=\mathcal V_1\left(1-\mathcal G_{11}\mathcal V_1\right)^{-1}$ and $T_2=\mathcal V_2\left(1-\mathcal G_{22}\mathcal V_2\right)^{-1}$ represent the scattering matrices for ${\rm X}_1$ and ${\rm X}_2$\cite{tmatrix}. Note that Eq. \eqref{uform} is derived without using any perturbative expansion, and thus it applies to strong scattering potentials. If the fermionic host is weakly correlated, $\mathcal G_{12}$ and $\mathcal G_{21}$ represent the renormalized Matsubara Green's functions with interactions encoded in their self-energy parts.  To the lowest order expansion, $\mathrm U_{12}=\frac{1}{\beta} \sum_{n}\mathcal G_{12}\, T_2\,\mathcal G_{21}\,T_1$ represents a two-fermion-exchange interaction, as illustrated in Fig. 1.  One could obtain the zero-temperature formula by replacing $\frac{1}{\beta}\sum_n$ with $\int \frac{d\xi}{2\pi}$ ($\xi$ is imaginary frequency) in Eq. \eqref{uform}, which agrees with the result in the literature\cite{levitov,lebo}. 

{\it Symmetry classification of Green's functions.---}We now present one more ingredient - the symmetry classification of Green's functions - aiming at a universal theorem for the sign of fermion-mediated interactions. In recent years, classification of Green's functions under  nonspatial symmetries, namely time-reversal symmetry,  charge-conjugation (particle-hole) symmetry and chiral symmetry, has been used to classify topological phases of correlated fermions\cite{gurarie}. For clarity, we start with the symmetry classification of non-interacting Hamiltonians\cite{ludwig} and then derive the symmetry classification of Matsubara Green's functions which turns out to be particularly useful when interactions are present. 

Consider a general system of non-interacting fermions described by the Hamiltonian
$
\hat H=\sum_{\alpha\beta}\psi^\dagger_\alpha \mathcal H_{\alpha\beta} \psi_\beta
$
with the fermion creation and annihilation operators satisfying canonical anti-commutation relations
$\{\psi_\alpha, \psi_\beta^\dagger \}=\delta_{\alpha\beta}$. Here, the indices $\alpha,\beta$ refer to relevant degrees of freedom, such as lattice sites, spins, layers, {\it etc.}; and $\mathcal H$, the first quantized Hamiltonian, is a complex matrix. Under the constraints of time-reversal symmetry and charge-conjugation symmetry, the Hamiltonian matrix $\mathcal H$ satisfies the following conditions:
\begin{subequations}
\begin{eqnarray}
\label{symmh1}U_T^\dagger \mathcal H^*(\mathbf k)U_T&=&\mathcal H(-\bold k)
\\
\label{symmh2}U_C^\dagger \mathcal H^*(\bold k)U_C&=&-\mathcal H(-\bold k)
\end{eqnarray}
\end{subequations}
where $U_T$ and $U_C$ are the corresponding unitary matrices. It is crucial to consider an additional discrete symmetry, the chiral symmetry, being the product of time-reversal symmetry and charge-conjugation symmetry.  The chiral symmetry imposes an additional condition on the Hamiltonian matrix
\begin{eqnarray}
\label{symmh3}U_S^\dagger \mathcal H(\bold k)U_S=-\mathcal H(\bold k).
\end{eqnarray}
with the chiral matrix $U_S=U_T^* U_C$. Notice that the chiral symmetry can be present in cases where neither time-reversal nor charge-conjugation symmetry is present. The above symmetry constraints lead to the ten-fold classification of topological insulators and superconductors\cite{ludwig}. 

Equipped with the above Hamiltonian formalism, we now proceed to present the symmetry classification of the Green's functions. We are interested in the Matsubara Green's functions defined as
$\mathcal G(i\omega_n, \bold k)=\left[i\omega_n-\mathcal H(\bold k)\right]^{-1}$. According to Eqs. \eqref{symmh1}, \eqref{symmh2}, and \eqref{symmh3}, time-reversal symmetry and charge-conjugation symmetry set the following conditions for Matsubara Green's functions, i.e.,
\begin{subequations}
\begin{eqnarray}
\label{utgreen} U_T  \mathcal G(i\omega_n,\bold k)U_T^\dagger=\frac{1}{i\omega_n-\mathcal H^*(-\bold k)}=\mathcal G^*(-i\omega_n,-\bold k),\nonumber\\
\\ \label{ucgreen} U_C  \mathcal G(i\omega_n,\bold k)U_C ^\dagger=\frac{1}{i\omega_n+\mathcal H^*(-\bold k)}=-\mathcal G^*(i\omega_n,-\bold k).\nonumber\\
\end{eqnarray}
\end{subequations}
Furthermore, chiral symmetry implies that the Matsubara Green's function should fulfill the condition
\begin{eqnarray}
\label{usgreen}U_S  \mathcal G(i\omega_n,\bold k) U_S^\dagger=-\mathcal G(-i\omega_n,\bold k)
\end{eqnarray}
consistent with Eqs. \eqref{utgreen} and \eqref{ucgreen}. Finally, regardless of whether interactions are present or not, the Hermicity of the Hamiltonian ensures that
\begin{eqnarray}
\mathcal G(i\omega_n,\bold k)=\mathcal G^\dagger(-i\omega_n, \bold k).
\end{eqnarray}
As a result, the combination of chiral symmetry and Hermicity leads to the momentum-space condition
\begin{eqnarray}\label{equsg}
U_S  \mathcal G(i\omega_n,\bold k) U_S^\dagger=-\mathcal G^\dagger(i\omega_n,\bold k),
\end{eqnarray}
which, after Fourier transform, yields the real-space expression
\begin{eqnarray}\label{equsgr}
U_S  \mathcal G(i\omega_n,\bold r) U_S^\dagger=-\mathcal G^\dagger(i\omega_n,-\bold r)
\end{eqnarray}
which is an essential ingredient in the proof of our theorem.
While the above Eqs. \eqref{utgreen}, \eqref{ucgreen}, and \eqref{usgreen} are obtained from non-interacting Hamiltonians, a sophisticated field-operator approach shows that the above Green's function formalism also holds for interacting systems. Interested readers could also find the proof from the excellent works Ref.\onlinecite{gurarie}. 

{\it The theorem.---}
In systems with symmetry, it is often convenient to choose a chiral basis such that the chiral operator $U_S$ is diagonal, i.e.,
$U_S={\rm diag}\left(\openone_n, -\openone_m\right)$,
where $\openone_{n}$ and $\openone_m$ are $n\times n$ and $m\times m$ identity matrices. The eigenvalue $+1$ ($-1$) denotes the chirality of the corresponding basis. One can then use the Pauli matrix $\tau_z$ to represent $U_S$ for each pair of basis states with opposite chiralities. (We use $\tau_{x,y,z}$ to represent Pauli matrices for general degrees of freedom, while reserving the notation $\sigma_{x,y,z}$ for real spin.) The Matsubara Green's function matrix in the pair chiral basis  $\left(|\chi \rangle,~|\bar\chi\rangle\right)$ reads
\begin{eqnarray}
\mathcal G(i\omega_n, \bold r)=
\left(\begin{array}{cc}
\mathcal G^{\chi\chi}(i\omega_n, \bold r)& \mathcal G^{\chi\bar\chi}(i\omega_n, \bold r)\\
\mathcal G^{\bar\chi\chi}(i\omega_n, \bold r)& \mathcal G^{\bar\chi\bar\chi}(i\omega_n,\bold r)
\end{array}\right),
\end{eqnarray}
where the indices $\chi=-\bar\chi=\pm 1$ represent chiralities of the corresponding basis. Apparently, the diagonal components of the Green's function connect objects with the same chirality, whereas the off-diagonal components connect objects with the opposite chiralities. 
By substituting the Matsubara Green's function and $U_S=\tau_z$ into Eq.\eqref{equsgr},
we obtain the crucial conditions required by the chiral symmetry:
\begin{subequations}
\begin{eqnarray}
\label{symmG1}&&\mathcal {G^{\chi\chi}}(i\omega_n,\bold r)=-\mathcal {G^{\chi\chi}}^*(i\omega_n,-\bold r),\\
\label{symmG2}&&\mathcal G^{\chi\bar\chi}(i\omega_n,\bold r)=\mathcal {G^{\bar\chi\chi}}^*(i\omega_n,-\bold r).
\end{eqnarray}
\end{subequations}
If two objects have the same chirality, the connecting Matsubara Green's functions satisfy $\mathcal G_{12}\equiv \mathcal G^{\chi\chi}(i\omega_n,\bold r)=-\mathcal G_{21}^*\equiv-\mathcal {G^{\chi\chi}}^*(i\omega_n,-\bold r)$. By contrast, if two objects have opposite chiralities, the connecting Matsubara Green's functions satisfy $\mathcal G_{12}\equiv \mathcal G^{\chi\bar\chi}(i\omega_n,\bold r)=\mathcal G_{21}^*  \equiv \mathcal {G^{\bar\chi\chi}}^*(i\omega_n,-\bold r)$.
As indicated by the Eq.\eqref{uform}, the sign of the fermion-mediated interaction is identical to the sign of the product $\mathcal G_{12} T_2\mathcal G_{21} T_1$\cite{footnote}. Due to the expression of the scattering matrices (below Eq. \eqref{uform}), two possibilities can be distinguished: 

\begin{enumerate}[label=\arabic*)]

\item{In the limit of a strong potential ($\mathcal V_i \rightarrow \infty$), the scattering matrices $T_1=T_2=\mathcal G^{-1}(i\omega_n,0)$ are purely imaginary according to the Eq. \eqref{symmG1}, and therefore $T_1 T_2<0$. It is then straightforward to obtain the sign of fermion-mediated interactions in the following cases: 
\begin{enumerate}[label=\roman*.]
\item{Between two objects with the {\it same} chirality, the fermion-mediated interactions are always {\it repulsive} because
\begin{eqnarray}
\begin{aligned}
{\rm sign\left(U_{12}\right)}&={\rm sign}\left(\mathcal G_{12}T_2 \mathcal G_{21}T_1\right)\\
&{\rm =-sign} \left(\mathcal G_{21}^*T_2 \mathcal G_{21}T_1\right)>0.
\end{aligned}
\end{eqnarray}}
\item{Between two objects with {\it opposite} chiralities, the fermion-mediated interactions are always {\it attractive} due to
\begin{eqnarray}
\begin{aligned}
{\rm sign\left(U_{12}\right)}&={\rm sign}\left(\mathcal G_{12}T_2 \mathcal G_{21}T_1\right)\\
&{\rm =sign} \left(\mathcal G_{21}^{*}T_2 \mathcal G_{21} T_1\right)<0.
\end{aligned}
\end{eqnarray}}
\end{enumerate}}
\item{In the limit of a weak potential ($\mathcal V_i \rightarrow 0$), the scattering matrices  $T_i=\mathcal V_i$ are purely real, and $T_1T_2>0$. Hence, the signs of fermion-mediated interactions are opposite to the case of strong potential: Between two objects with the {\it same} ({\it opposite}) chirality, the fermion-mediated interactions are always {\it attractive} ({\it repulsive}). }
\end{enumerate}
As the above results can be conveniently summarized in Eq.\eqref{signu}, we have thus established the theorem for the sign of fermion-mediated interactions. The above analysis indicates that one could change the sign of fermion-mediated interactions by tuning the strength of scattering potentials. To demonstrate the power of this theorem, we now explore a number of examples with an emphasis on electron-mediated interactions in materials.

{\it Electron-mediated interactions in non-interacting systems.}
As our first example, let us consider electron-mediated interactions between two adatoms embedded in monolayer graphene.
The physics of single layer graphene is captured by the continuum low-energy Hamiltonian $\mathcal H_{\text{SLG}}=v_F\left(k_x\tau_x+k_y\tau_y\right)$, where $\tau_{x,y,z}$ represent the Pauli matrices in the sublattice basis, and $v_F$ is the Fermi velocity\cite{castro}.
One can easily identify the chiral transformation matrix $U_S=\tau_z$. Therefore, without further detailed calculation, we can apply the theorem to graphene. In the strong-impurity limit\cite{wehling}, adatoms residing on the same (different) sublattices repel (attract), whereas, in the weak-impurity limit, impurities reside on the same (different) sublattices attract (repel). This is consistent with detailed calculations carried out in Refs.\cite{levitov,lebo} One can consider more general Hamiltonians such as $\mathcal H=v_F\left(k_x^n\tau_x+k_y^m\tau_y\right)$, where $m$ and $n$ can be arbitrary odd numbers. As the chiral matrix for this Hamiltonian is still $U_S=\tau_z$, our theorem also applies to this model. 

Our second example deals with the electron-mediated interactions in the Bernal-stacked bilayer graphene, the unit cell of which includes $\rm A_1$ and $\rm B_1$ atoms on layer 1 and $\rm A_2$ and $\rm B_2$ atoms on layer 2. Expressed in the basis $(\psi_{A1},\psi_{A2},\psi_{B1},\psi_{B2})$ (here subindices denote the lattice type), the low-energy effective Hamiltonian reads\cite{blgchiral}
\begin{eqnarray}\label{blghamiltonian}
\mathcal H_{\text{BLG}}=\left(
\begin{array}{cc}
0&\mathcal D^*(-\bold r)\\
\mathcal D(\bold r)&0
\end{array}\right), \,
\mathcal D(\bold r)=\left(
\begin{array}{cc}
-i\partial_+&\gamma\\
0&- i \partial_+
\end{array}\right),\nonumber\\
\end{eqnarray}
where $\partial_{\pm}=\partial_x\pm i \partial_y$ and $\gamma$ denotes the interlayer coupling; we set the fermi velocity $v_F=1$.
It can be verified that this Hamiltonian preserves chiral symmetry with the chiral matrix $U_S=\tau_z\otimes \openone$,
which indicates that sublattices ($\rm A_1$, $\rm A_2$) have positive chirality while sublattices ($\rm B_1$, $\rm B_2$) have negative chirality.  Accordingly, regardless the layer index, the electron-mediated interactions are always repulsive (attractive) between impurities on the same type of sublattices (i.e., AA or BB) in the strong (weak) impurity limit. In contrast, when impurities reside on different sublattices (i.e., AB), the electron-mediated interactions are always attractive (repulsive) in the strong (weak) impurity limit. 

One may also apply the theorem to study the twisted bilayer model, which has generated considerable interest recently.  A specific chiral-symmetric model\cite{vish} that supports exactly flat bands has a similar form as Eq. \eqref{blghamiltonian} but with a different $\mathcal D$ matrix
\begin{eqnarray}
\mathcal D_{\text{tBLG}}(\bold r)=\left(
\begin{array}{cc}
-v_F i\partial_+&\mathcal F(\theta, \bold r)\\
\mathcal F(\theta, -\bold r)&-v_F i \partial_+\\
\end{array}\right),
\end{eqnarray}
where the moir\'e potential, $\mathcal F(\theta, \bold r)$, is a function of the twist angle $\theta$ and position $\bold r$.  Regardless of the specific form of $\mathcal F(\theta, \bold r)$, this model preserves chiral symmetry with the same chiral matrix as Bernal-stacked bilayer graphene. Consequently, the sign of electron-mediated interaction in twisted bilayer model shares the identical feature with that in the Bernal-stacked bilayer graphene. The chiral-symmetry protected sign of fermion-mediated interactions could be of essential importance for understanding disorder-related physics in {twisted bilayer models.} We should, however, clarify that twisted bilayer materials in experiments may violate chiral symmetry due to substrate or interactions.

{\it Electron-mediated interaction in correlated systems.---}
Our theorem is applicable to weakly interacting systems. We examine electron-mediated interactions in the Hubbard model defined on a bipartite lattice
\begin{eqnarray}
\hat H=\sum_{\langle i,j\rangle,\sigma }t_{ij}\hat c_{i,\sigma}^\dagger \hat c_{j,\sigma}-\mu\sum_{i}\hat n_{i\sigma}+U\sum_i\hat n_{i\uparrow}\hat n_{i\downarrow},
\end{eqnarray}
where $\hat c_{i,\sigma}$ and $\hat c_{i,\sigma}^\dagger$ is the electron annihilation and creation operators at site $i$ with spin $\sigma=\uparrow$ or $\downarrow$ and $\hat n_{i\sigma}=\hat c_{i\sigma}^\dagger \hat c_{i\sigma}$. Here, the nearest-neighbor hopping matrix elements $t_{ij}=t_{ji}$ need to be real; $\mu$ and $U$ represent the chemical potential and on-site interaction strength, respectively. One can verify that the Hamiltonian is invariant at half-filling ($\mu=U/2$)
under the following chiral symmetry transformation: $\mathcal{\hat S} \hat c_{i\uparrow/\downarrow} \mathcal{\hat S}^{-1}=\kappa(i)~ \hat c_{i\downarrow(\uparrow)}^\dagger$,  $\mathcal{\hat S} \hat c_{i\uparrow(\downarrow)}^\dagger \mathcal{\hat S}^{-1}=\kappa(i)~ \hat c_{i\downarrow(\uparrow)}$ with $\kappa(i)=1$ for one of the sublattices and $\kappa(i)=-1$ for the other.  It should be obvious that the same sublattices have the same chirality and different sublattices have the opposite chiralities.  We assume that $U$ is small enough for the ground state to preserve chiral symmetry so that our result Eq.\eqref{uform} still holds. Without performing any perturbative calculation, we can apply our theorem to the correlated bipartite lattices: fermion-mediated interactions are repulsive or attractive depending on the impurities located on the same or different sublattices. Our theorem can be applied to all weakly coupled models with chiral symmetry, such as BdG systems with time-reversal symmetry\cite{bdgchiral} and QCD at high density\cite{qcd}.

{\it RKKY interactions in systems with chiral symmetry.---}A similar line of reasoning enables us to predict the sign of RKKY interactions in the presence of chiral symmetry: magnetic moments with the {\it same} ({\it opposite}) chirality favor a {\it ferromagnetic} ({\it antiferromagnetic}) state. We demonstrate this statement by considering two magnetic moments ($\bold S_1$ and $\bold S_2$) embedded in a fermionic system described by the effective action
\begin{eqnarray}
\mathcal S =-\sum_{n}\int && d\bold x \, \bar\psi_n(\bold x)\left[ \mathcal G^{-1}-\frac{J}{2}\bold S_{1}\cdot \boldsymbol\sigma\delta(\bold x-\bold x_1)\right.\nonumber\\
&&\left.-\frac{J}{2} \bold S_{2}\cdot \boldsymbol\sigma \delta(\bold x-\bold x_2)\right]\psi_n(\bold x).
\end{eqnarray}
with $\mathcal G$ being the Green's function of the host system and $J$ as the coupling constant between the impurity magnetic moment and the spin of electrons $\sigma_{\mu}$ ($\mu=x,y,z$)  located at $\bold x_{1}$ and $\bold x_2$, respectively. Note that this action has the same form as Eq. \eqref{actions} given the substitution $\hat{\mathcal V}_i=\bold S_i\cdot \boldsymbol \sigma\delta(\bold x-\bold x_i)$.  According to Eq. \eqref{uform}, we then obtain the interaction energy of the nuclear spins (to the lowest order of $J$):
\begin{eqnarray}\label{chiralrkky}
{\rm U_{12}}&=&\frac{J^2}{4\beta} \sum_{\mu,\nu,n}{\rm  tr}\left[S_{1\mu} \sigma_\mu \mathcal G(i\omega_n,\bold r)S_{2\nu}\sigma_\nu \mathcal G(i\omega_n,-\bold r)\right]\nonumber\\
&=&-J^2\alpha(\bold r)\,\bold S_1\cdot \bold S_2,
\end{eqnarray}
where the susceptibility is defined as 
$\alpha(\bold r)=-\frac{1}{2\beta}\sum_{n} \mathcal G (i\omega_n,\bold r) \mathcal G (i\omega_n,-\bold r)$ with $\bold r=\bold x_2-\bold x_1$ as the relative distance between the localized spins. 
Note that we have assumed that the Green's function is spin-independent in the above derivation. Additional term such as Dzyaloshinskii-Moriya type of interaction can emerge in spin-orbital coupled systems where the Green's functions depend on spin\cite{DMinteraction,simon}. 
It is also desirable to generalize Eq. \eqref{chiralrkky} to SU(N) RKKY interactions due to the remarkable progress in experiments\cite{kondo,append}.

The two possibilities, $\alpha(\bold r)>0$ ($\alpha(\bold r)<0$) corresponds to a ferromagnetic (antiferromagnetic) alignment of magnetic moments.
Consider two magnetic moments (spins)  that are embedded in a system with chiral symmetry,  and have definite chiralities $\chi_1$ and $\chi_2$, respectively. The susceptibility can then be written as
\begin{eqnarray}
\alpha^{\chi_1\chi_2}=-\frac{1}{2\beta}\sum_{n} \mathcal G^{\chi_1\chi_2}(i\omega_n,\bold r) \mathcal G^{\chi_2\chi_1}(i\omega_n,-\bold r). 
\end{eqnarray}
Based on the symmetry properties of Green's functions (see Eqs. \eqref{symmG1} and \eqref{symmG2}), we can verify that
$\alpha^{\chi \chi}>0$ and $\alpha^{\chi \bar\chi}<0$. As a result, magnetic moments with the same (opposite) chirality favor a ferromagnetic (antiferromagnetic) alignment. This applies to all the examples discussed in the previous sections.
Note that our theorem exhibits two major differences compared to the theorem concerning the sign of RKKY interactions on noninteracting bipartite lattices\cite{saremi,kogan}. First, our theorem has a broader scope of application, as the bipartite model is only a specific system preserving chiral symmetry. Second, our theorem can be applied to interacting models as long as chiral symmetry is present.  
This indicates the possibility of chiral symmetry breaking in the previous calculations, which showed that the sign of the RKKY interactions can be modified by electronic interactions, edge states, strains, or flat bands\cite{blackschaffer}. 

{\it Concluding Remarks.---}We have established a universal theorem regarding the sign of fermion-mediated interactions in systems with chiral symmetry. We have demonstrated the strength of the theorem by considering multiple examples ranging from noninteracting models to weakly correlated systems. Without involving any spatial symmetry, the theorem is robust to disorders and defects. Furthermore, our theorem suggests a new route to probe chiral symmetry breaking via fermion-mediated interactions: When additional parameters break chiral symmetry, the sign of fermion-mediated interactions will not be definite, and will oscillate with respect to these parameters. In the future, it will be interesting to investigate how other symmetries ({\it e.g.} crystalline symmetries) constrain fermion-mediated interactions.
It would also be desirable to study the sign of two-boson-mediated interactions in systems with chiral symmetry\cite{jiang2}.

{\it Acknowledgements.---}We thank F. Wilczek for his encouragement and support, and E. Bergholtz, L. Liang, Y. Kedem, and especially T. H. Hansson for indispensable clarifying discussions. We thank K. Dunnett for her proofreading of the manuscript. This work was supported by the Swedish Research Council under Contract No. 335-2014-7424.


\appendix
\begin{widetext}
\begin{center}
\textbf{Supplemental Material for ``On the sign of fermion-mediated interactions"}
\end{center}

Here, we discuss a possible application of the theorem to study SU(N) RKKY interactions as a generalization of the SU(2) RKKY interactions ( i.e., Eq. (21)) in the main text. This is motivated by the remarkable progress in experiments in recent years\cite{kondo}. Such a generalization could be achieved by replacing the role of the SU(2) electron spin with an SU(N) pseudospin, which can be realized by using other quantum degrees of freedom (e.g. orbital momentum)\cite{kondo2}. Tracing out the SU(N) fermionic degrees of freedom, one may obtain the SU(N) RKKY interactions. To do that, let us consider two impurities embedded in a SU(N) Hubbard model, which is captured by the following Hamiltonian 
\begin{eqnarray}
\hat H=\sum_{\langle i,j\rangle, \sigma} t_{ij} \hat c_{i,\sigma} ^\dagger\hat c_{j,\sigma} -\mu\sum_{i} \hat n_{i,\sigma}+U\sum_{i,\sigma\neq \sigma^\prime} \hat n_{i,\sigma}\hat n_{i,\sigma^\prime}+ \frac{J}{N}\sum_{i=1,2}\sum_{\mu=1}^{N^2-1} {\hat \Gamma}_{\mu}^c \, S_{i\mu}^N.
\end{eqnarray}
Here, $\sigma$ is a SU(N) spin index; ${\hat \Gamma_\mu^c}=\hat c^\dagger \Gamma_\mu \hat c$ represent the SU(N) spin operators for conduction electrons with $\Gamma_\mu$ being the $N^2-1$ generators of SU(N) group; $ S_{i\mu}^N$ denotes the spin of the impurity at $\bold x_i$. The definitions of other parameters are the same as in the main text. While the first three terms describe the SU(N) extension of a Hubbard model on a bipartite lattice, the last term describes the spin coupling between conduction electrons and embedded impurities. 
Substitute the coupling term into the expression of fermion-mediated interactions, and we can obtain a similar formula as Eq.(21) with the replacements $S_\mu\rightarrow S_\mu^N$ and $J\rightarrow J/N$. Notice that we should use the normalization condition ${\rm Tr}\left(\Gamma_\mu\Gamma_\nu\right)=\frac{\delta_{\mu\nu}}{2}$ in the derivation; meanwhile,  we should assume that the Green's functions are independent of SU(N) pseudo spins. If the ground state preserves the chiral symmetry, one could apply our theorem to determine the sign of this SU(N) RKKY interaction. Nevertheless, one should be aware that a chiral-symmetric ground state may be difficult to realize in a general SU(N) spin model\cite{hermele}.
\end{widetext}

%
%
%
\end{document}